\documentclass[prb,superscriptaddress, preprintnumbers ,twocolumn,showpacs,amsmath,amssymb]{revtex4}
\usepackage{amsmath}
\usepackage{amssymb}
\usepackage{amsthm}
\usepackage{amsfonts}
\usepackage{enumerate}
\usepackage{latexsym}
\usepackage{graphicx}

\newcommand{\be}{\begin{equation}}
\newcommand{\ee}{\end{equation}}
\newcommand{\ba}{\begin{eqnarray}}
\newcommand{\ea}{\end{eqnarray}}
\newcommand{\baa}{\begin{eqnarray*}}
\newcommand{\eaa}{\end{eqnarray*}}

\def\be{\begin{equation}}
\def\ee{\end{equation}}
\def\ba{\begin{eqnarray}}
\def\ea{\end{eqnarray}}

\def\C60{A$_x$C$_{60}$}

\def\HgCu3{HgCa$_2$Cu$_3$O$_{8+y}$}
\def\HgCu4{HgBa$_2$Ca$_3$Cu$_4$O$_{10+y}$}
\def\TlCu{Tl$_2$Ba$_2$CuO$_{6+\delta}$}
\def\TlCu3{Tl$_2$Ba$_2$Ca$_2$Cu$_3$O$_{10+y}$}
\def\TlCu4{Tl$_2$Ba$_2$Ca$_3$Cu$_4$O$_{12+y}$}

\def\BiCu3{Bi$_2$Sr$_2$Ca$_{2}$Cu$_3$O$_y$}

\def\8LSCO{La$_{1.88}$Sr$_{.12}$CuO$_4$}
\def\110LNSCO{La$_{1.5}$Nd$_{0.4}$Sr$_{0.1}$CuO$_{4}$}
\def\stage4LCO{La$_{2}$CuO$_{4+\delta}$}
\def\Y248{YBa$_2$Cu$_4$O$_8$}

\def\NbSe2{NbSe$_2$}
\def\TaSe2{TaSe$_2$}
\def\TiSe2{TiSe$_2$}

\begin{document}
\title{ Theory of Magnetic Order in  $ Fe_{1+y}Te_{1-x}Se_x$}
\author{Chen Fang}
%\email{ seo@physics.purdue.edu }
\affiliation{Department of Physics, Purdue University, West Lafayette, Indiana 47907, USA}

\author{B. Andrei Bernevig}
\affiliation{Princeton Center for Theoretical Science, Princeton
University, Princeton, NJ 08544}

\author{Jiangping Hu}
%\email{ hu4@physics.purdue.edu}
\affiliation{Department of Physics, Purdue University, West Lafayette, Indiana 47907, USA}

\begin{abstract}
We develop a local spin model to explain the rich magnetic
structures in the iron-based superconductors $Fe_{1+y}Te_{1-x}Se_x$.
We show that our model exhibits both commensurate antiferromagnetic
and incommensurate magnetic order along the crystal a-axis. The
transition from the commensurate to the incommensurate phase is
induced when the concentration of excess $Fe$ atoms is larger than a
critical value. Experimentally measurable spin-wave features are
calculated, and the mean-field phase diagram of the model is
obtained. Our model also suggests the existence of a large quantum
critical region due to strong spin frustration upon increasing $Se$
concentration.
\end{abstract}
\date{\today}
\maketitle

%{\it Introduction:}
 Superconductivity with critical temperature
$T_c=14K$ was recently reported in the iron-selenide-telluride
compound $Fe_{1+y}Te_{1-x}Se_x$
\cite{Hsu2008,Yeh2008,Fang2008se,Margadonna2008} with $0\le x \le
1$. This discovery not only adds a new class of iron-based
superconductors to the multitude of already existing ones, but also
provides a fresh angle to investigate the fundamental physics of the
$Fe-As$ based
superconductors\cite{kamihara2008,Chenxh2008,Chen2008,wen2008}.
$Fe_{1+y}Te_{1-x}Se_x$, similar to the $Fe-As$ based materials, has
a PbO structure of square planar
  sheets of tetrahedrally coordinated $Fe$ atoms. The electronic band structure of $Fe_{1+y}Te_{1-x}Se_x$ calculated by LDA is
  very close to that of the $Fe-As$ based superconductors\cite{Subedi2008c}. It exhibits electron pockets at the
zone corner and hole pockets at the zone center. Similar to the
$Fe-As$ based superconductors, it is believed that the magnetism in
$Fe_{1+y}Te_{1-x}Se_x$ plays an important role in forming electron
Cooper pairs.

The  parent compounds of the $Fe-As$ based superconductors exhibit
stripe-type commensurate antiferromagnetic spin
order\cite{Cruz2008}. However, the origin of the spin order has been
theoretically controversial, with two very different mechanisms
leading to the same physical answer. One theory is based on Fermi
surface nesting between the electron and hole pockets at the zone
corner and center respectively\cite{Mazin2008a}. This weak-coupling
approach leads to a commensurate spin density wave (SDW) state at
the nesting wavevector, as observed in the experiments on $Fe-As$
based parent materials.  The competing view is that, due to the
geometry of As-mediated hopping, antiferromagnetic exchange exists
not only between the nearest neighbor (NN) $Fe$ sites, but also
between next nearest neighbor (NNN) sites\cite{Yildirim2008,Ma,si,
Fang2008,seo2008}. Moreover, the NNN coupling strength $J_2$ is
stronger than  the half of the NN coupling strength $J_1/2$. The
resulting $J_1 - J_2$ model produces magnetic physics consistent
with the experimental results. Although they lead to the same
overall prediction, the two theories rely on different mechanisms to
describe the parent state's magnetism, and a conclusive test of
either of them is needed.

\begin{figure}[t]\vspace{-0pt}
\includegraphics[width=8cm, height=6cm]{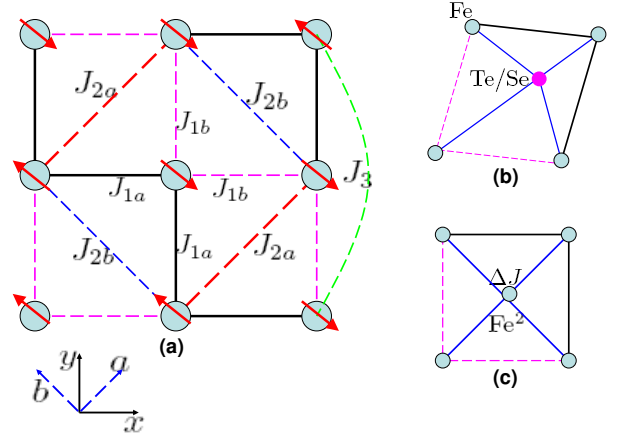}
\caption{(a) $Fe$ (noted as $\text{Fe}^1$) square lattice showing
commensurate AFM3 order along the a-axis. The magnetic exchange
parameters $J_{1a},J_{1b},J_{2a}$ and $J_{2b}$ are specified. (b)
The monoclinic lattice distortion observed in neutron
scattering\cite{Bao2008,Li2008}. (c) The magnetic exchange coupling
between $\text{Fe}^1$ and an excess Iron atom in the middle of the
plaquette indicated as $\text{Fe}^2$.
   \label{lattice}}
\end{figure}

We believe that the recent neutron scattering data in
$Fe_{1+y}Te_{1-x}Se_x$\cite{Bao2008,Li2008} sheds new light on the
origin of magnetism in $Fe$-based materials. A commensurate spin
order along the $a$ axis (see Fig[\ref{lattice}]) in $FeTe$ was
reported. This gives way to an incommensurate spin order along the
same axis upon the introduction of excess $Fe$ atoms. The
commensurate spin order in $FeTe$ is different from the one in the
$Fe-As$ parent compounds: the two ordered wavevectors are rotated by
$45$ degrees with respect to each other. This experimental fact
places a clear challenge to theories based on Fermi surface nesting.
The Fermi surfaces of $Fe_{1+y}Te_{1-x}Se_x$ are predicted, from LDA
studies, to be similar to those in the $Fe-As$ based materials: the
electron and hole pockets are separated by a 2D nesting vector at
$(\pi,0)$. Based on first-principle calculations, it was  then
predicted that $FeTe$ should support an identical spin ground state
to that observed in $Fe-As$ materials\cite{Subedi2008c}. With the
experiment falsifying this prediction, the spin order in $FeTe$
cannot be, at least trivially, understood by a Fermi surface nesting
mechanisms.

In this letter, we show that the magnetic physics in the $FeTe$
parent compound can be
 understood from the usual magnetic exchange nearest and next nearest neighbor $J_1-J_2$ model used for the $Fe-As$ based materials, but with a natural
 parameter extension that takes into account the monoclinic lattice
 distortion observed in these compounds.
  The lattice distortion in $ Fe_{1+y}Te_{1-x}Se_x$ is
 different from the one in $Fe-As$ based materials. The two lattice distortion directions form a $45$-degree angle, just like the magnetic wavevectors in the magnetic ordering states of these two systems.
  Our extended $J_1-J_2$ model can explain both the commensurate and the incommensurate
 spin order phases along the a-axis which have been measured in neutron scattering experiments.
 The commensurate to incommensurate phase transition takes place at
a critical concentration of excess $Fe$ atom in
$Fe_{1+y}Te_{1-x}Se_x$. Above the critical concentration, the
incommensurate wavevector is proportional to the square root of the
concentration difference of excess $Fe$ atoms.

%{\it Model:}
We start with the $J_1-J_2$ model on the tetragonal lattice. Due to
their proximity in temperature,  we strongly believe that the
lattice and magnetic transitions in $Fe$-based materials are
physically related. Considering the coupling between the lattice and
magnetism, it is physical to assume that a particular lattice
distortion favors changes in the values of $J_1$ and $J_2$ as
follows:  in the $Fe-As$  compounds, $J_1$ should be slightly more
sensitive to \emph{changes} in the angle of the $Fe-As-Fe$
  bond (As is out of plane) than $J_2$. This is because the angle between two
  nearest $Fe$ atoms is around 72 degrees and hence closer to 90 degrees than the one between two next nearest
  $Fe$ atoms which is around 112 degrees. However, in
  $ Fe_{1+y}Te_{1-x}Se_x$, $J_2$ should be significantly more sensitive than $J_1$ to
  \emph{changes} in the angle between two nearest $Fe$ atoms; the angle between  two nearest $Fe$ atoms,
  which influences $J_1$, is around 66 degrees, whereas the angle between the two next nearest
  $Fe$ atoms is around 96 degrees and hence much closer to 90 degrees.

In the monoclinic lattice distorted phase,  the extended magnetic
Hamiltonian can be written as an in-plane
nearest and next nearest neighbor  Heisenberg model supplemented by
an out-of plane small antiferromagnetic coupling as well as an next
next nearest neighbor term:
\begin{eqnarray}& H=J_z \sum_{i,n} \vec S_i^n \vec
S_i^{n+1} +\sum_n \sum_{<ij>}J_{ij}\vec S_i^n \vec S_j^n +
\nonumber \\ &+ J_3 \sum_n \sum_{<<ij>>} \vec{S}_i^n \vec{S}_j^n,
\end{eqnarray} \noindent where $n$ is the layer index. The
$J_{ij}$'s, defined in Fig.[\ref{lattice}], are the magnetic exchange
coupling parameters between irons in the $a-b$ plane, and their
values depend on the lattice distortion direction.
 If small, the added $J_3$ next next nearest neighbor coupling, suggested by first-principle calculations \cite{Ma2008},
  influences the phase diagram only quantitatively. We take $J_{2a}\geq J_{2b}$,
$J_{1a}\geq J_{1b}$ and study the part of the phase diagram of the
model for which $(J_{2a},J_{2b})>0$ and $J_{1a}>0$. These values are
naturally expected in $ Fe_{1+y}Te_{1-x}Se_x$, as shown in Fig.[\ref{lattice}].

\begin{figure}[t]\vspace{-25pt}
\includegraphics[width=8cm, height=9cm]{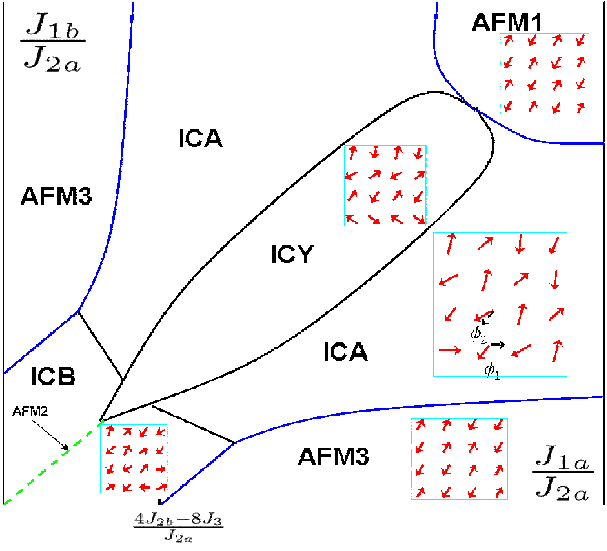}
\caption{The phase diagram of the
$J_1(J_{1a},J_{1b})-J_2(J_{2a},J_{2b}) -J_3$ model: (1) AFM1 is the
familiar $(\pi,\pi)$ antiferromagnetic order in a square lattice;
(2) AFM2 is the $(0,\pi)$ collinear AFM order along the $x$-axis
observed in the $Fe-As$ based superconductors; it is possible in the
present model when $J_{1a}=J_{1b}$, indicated as the green line in
the figure; (3) AFM3 is the collinear commensurate  AFM order along
the $a$-axis observed in $ Fe_{1+y}Te_{1-x}Se_x$;(4) ICB is an
incommensurate phase with the incommensurate vector along the
b-axis; (5) ICA is an incommensurate phase with the incommensurate
vector along the a-axis which has been observed in $
Fe_{1+y}Te_{1-x}Se_x$\cite{Bao2008}; (6) ICY is an incommensurate
phase that has a wavevector of the form $(\pi,q)$.
   \label{phase}}
\end{figure}

%{\it Phase Diagram:}
As the exchange along the c-axis $J_z>0$ is not frustrated, we only
focus on in-plane magnetism. All the possible ground states are
presumed to have a $k_z=\pi$.  The classical ground state of the
Hamiltonian can be obtained by comparing the energy of the following
six states:(1) $(\pi,\pi)$ antiferromagnetic (AFM) phase (AFM1
phase) with $E_1=-J_{1a}-J_{1b}+J_{2a}+J_{2b}+2J_3$;  (2 ) $(0,\pi)$
AFM state (AFM2 phase) with $E_2=-J_{2a}-J_{2b}+2J_3$; (3)
Commensurate AFM along the $a$-axis(AFM3) with $E_3=
-J_{1a}-J_{2a}+J_{1b}+J_{2b}-2J_3$   (see Fig(\ref{lattice})); (4)
Incommensurate AFM along the $b$-axis(ICB phase) with energy
$E_4=Min(-J_{2a}+(J_{1a}-J_{1b})cos(\phi)+(J_{2b}+2J_3)cos(2\phi))$,
to be minimized over $\phi$; (5) Incommensurate AFM along the
$a$-axis (ICA phase) with $ E_5= Min(J_{2b}+J_{1a}
cos(\phi_1)+J_{1b}cos(\phi_2)+ (J_{2a}+2J_3)cos(\phi_1+\phi_2)) $,
to be minimized with respect to $\phi_1$ and $\phi_2$; (6)
Incommensurate AFM along the $y$-axis (ICY phase), with $
E_6=Min(-\frac{J_{1a}+J_{1b}}{2}+(J_{2a}+J_{2b}-\frac{J_{1b}+J_{1a}}{2})cos(\phi_3)+J_3
cos(2\phi_3)+J_3) $, to be minimized over $\phi_3$. The three
incommensurate phases are depicted in Fig.[\ref{phase}].

The phase diagram of above model in the
$\frac{J_{1a}}{J_{2a}}-\frac{J_{1b}}{J_{2a}}$ plane
 is plotted in Fig.[\ref{phase}].
For $FeTe$, we are interested in the AFM3 and ICA phases, which have
been experimentally observed. The AFM3 phase exists when the
following two conditions are satisfied: $J_{1a}\geq
J_{1b}+4J_{2b}-8J_3$ and $\frac{J_{1b}}{J^t_{2a}}\leq
\frac{J_{1a}}{J_{1a}+J^t_{2a}}$ where $J_{2a}^t= J_{2a} + 2J_3$. The
transition line between AFM3 and ICA phases is determined by
equality in the latter condition. In the ICA phase, the spin angle
difference of two next nearest neighbours $Fe$ atoms $\phi_1$ and
$\phi_2$, indicated in Fig.\ref{phase}  are given by \ba
\cos(\phi_{(1,2)})=\mp\frac{J_{2a}^t(J_{1a}^2-J_{1b}^2\pm\frac{J_{1a}^2J_{1b}^2}{(J^t_{2a})^2})}{2J_{1(a,b)}^2J_{1(b,a)}}.
 \ea The incommensurate wavevector, $\delta$ along the a-axis
corresponding to the definition in \cite{Bao2008} is
$\delta=\frac{\phi_1+\phi_2}{2\pi}$. If the next next nearest
neighbor coupling $J_3$ is large, its effect on the phase
diagram is qualitative. It can be analytically shown that if
$J_3>J_{2b}/2$, then the ICB phase disappears; while if
$J_3>J_{2a}/2$, then the ICY phase vanishes.

%{\it Commensurate to incommensurate transition:}
We now phenomenologically justify the parameters $J_{1a},J_{1b},
J_{2a},J_{2b}$ and their capacity to induce a
commensurate-incommensurate magnetic order transition upon doping
with excess $Fe$ atoms denoted as $\text{Fe}^2$ (experimentally
introduced in $FeTe$). The excess $\text{Fe}^2$ atoms are located at
the mirror-symmetric site of the $Te$ atom with respect to the $Fe$
layer.   This suggests that the additional $Fe^2$ directly couples
to the four nearest neighbor $Fe$  as shown in Fig.[\ref{lattice}c]
with coupling parameter $\Delta J$. In the AFM3 phase, regardless of
the sign of $\Delta J$, the excess $Fe^2$ spins align along the
b-axis through the $\Delta J$ coupling with the collinear spins
along the b-axis. The $\Delta J$ coupling along the $a$-axis is then
frustrated. The effective result of $\Delta J$ on the magnetic model
can be described by changing the effective parameters $\tilde
J_{1a}$, $\tilde J_{1b}$ and $\tilde J_{2a}$ as $\tilde J_{1a}(y)=
J_{1a}+y \Delta J, \tilde J_{1b}(y)= J_{1b}+y \Delta J$ and $\tilde
J_{2a}(y)= J_{2a}+y^2 \Delta J,$ where $y$ is the density of the
excess $Fe$ atoms. The change of $J_{2a}$ is a second order in $y$,
which can be ignored for low concentration of $Fe^2$. According to
the phase diagram in Fig.[\ref{phase}], the excess $Fe$ atoms can
cause a commensurate-incommensurate transition if $\Delta J$ is
antiferromagnetic, i.e. positive. From the phase boundary between
AFM3 to ICA we can then determine the critical concentration of
$Fe^2$ that would cause the transition from AFM3 to ICA: \ba y_c=
\frac{\sqrt{(J_{1a}-J_{1b})(J_{1a}+4J^t_{2a}-J_{1b})}-(J_{1a}+J_{1b})}{\Delta
J} \ea Away from the critical concentration, the incommensurate
angles as a function $\delta y=y-y_c$ are given by \ba
 \cos(\phi_1)&=&-1+\frac{2J^t_{2a}+J_{1a}}{J^t_{2a}+J_{1a}} \delta y\Delta J \\
\cos(\phi_2)&=&1-\frac{2(J^t_{2a})^2+3J_{1a}J^t_{2a}+J_{1a}^2}{J^t_{2a}
J_{1a}}\delta y \Delta J\ea Therefore the incommensurate
wavevector close to the transition is given by  \ba \delta= \frac{1}{2}-\frac{1}{2\pi}
(\frac{J_{1a}+J_{2a}^t}{\sqrt{J_{1a}J_{2a}^t}}-1)\sqrt{\frac{2(J_{1a}+2J_{2a}^t)\Delta
J \delta y}{J_{2a}^t(J_{1a}+J_{2a}^t)}} \ea

%(\sqrt{\frac{2(J^t_{2a})^2+3J_{1a}J^t_{2a}+J_{1a}^2}{J^t_{2a}
%J_{1a}}}-\sqrt{\frac{2J^t_{2a}+J_{1a}}{J^t_{2a}+J_{1a}}})
%\sqrt{2\frac{\Delta J}{J^t_{2a}}\delta y}\ea

\begin{figure}[t]
\includegraphics[width=8.5cm, height=6.5cm]{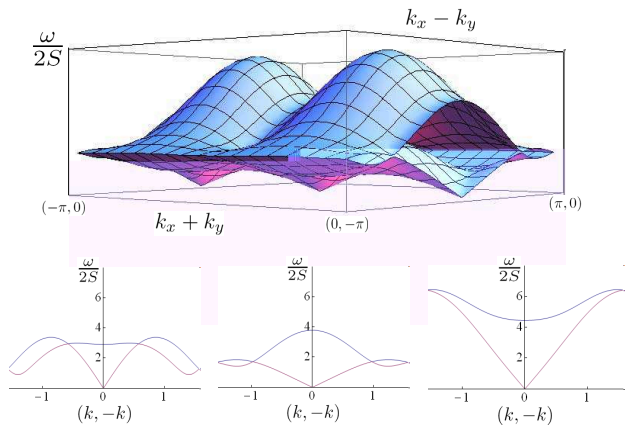}
\caption{3D spinwave dispersion for the parameter set of
$J_{2a}=J_{2b}=1$, $J_{1a}=2$, $J_{1b}=0.94$ and $J_3=0.4$, which
gives a point on the AFM3-ICA phase boundary on the phase diagram.
Also note that along the BZ boundary $(k\pm\pi,k)$, the two sheets
are degenerate and almost dispersionless. Below: Different
dispersion at different parameter regions with common parameters
$J_{2a}=J_{2b}=1$. All the three dispersions are drawn at AFM3-ICA
phase boundary given by (from left to right)
$(J_{1a},J_{1b},J_3)=(1,0.72,0.8)$,
$(J_{1a},J_{1b},J_3)=(2,0.94,0.4)$ and
$(J_{1a},J_{1b},J_3)=(5,0.9,0.1)$ respectively.
   \label{spinwave}}
\end{figure}

%
%{\it Spin wave and large $S$ limit: }
To predict experimentally observable quantities, we perform a
spinwave analysis on the AFM3 state. Suppose that the system is in
the classical AFM3 ground state, in which the spins are aligned as
in Fig.[\ref{lattice}]. This is the commensurate order recently
observed in the neutron scattering experiments. By introducing the
standard Holstein-Primakoff bosons
%$b_{ir}, b_{ir}^+$
%$S_{ir}^+=\sqrt{2S-b^\dag_{ir}b_{ir}}b_{ir}$,
%$S^-_{ir}=\sqrt{2S-b^+_{ir}b_{ir}}b^\dag_{ir}$,
%$S^z_{ir}=S-b^\dag_{ir}b_{ir}$ for the spins that are 'down' in the
%ground state, and $S_{ir}^+=\sqrt{2S-b^\dag_{ir}b_{ir}}b^\dag_{ir}$,
%$S^-_{ir}=\sqrt{2S-b^\dag_{ir}b_{ir}}b_{ir}$,
%$S^z_{ir}=-S+b^\dag_{ir}b_{ir}$ for spins that are 'down' in the
%ground state. The first subscript indicates the position of the
%spins in the lattice while the second their position in a unit cell.
%Keeping the terms that are quadratic in boson operators only,
we obtain the spin wave excitations,
\ba\label{spinwavehamiltonian}H=2S\sum_{i,j;k}[A_{i,j;k}b^+_{i,k}b_{j,k}+\frac{B_{i,j;k}}{2}(b^\dag_{i,k}b^\dag_{j,-k}+b_{i,-k}b_{j,k})],\ea
where\baa\label{}A_k=\left(%
\begin{array}{cc}
  \epsilon_1(k) & J_{1b}f(k) \\
  J_{1b}f(k)^* & \epsilon_1(k) \\
\end{array}%
\right), B_k=\left(%
\begin{array}{cc}
  \epsilon_2(k) & J_{1a}f(k)^* \\
  J_{1a}f(k) & \epsilon_2(k) \\
\end{array}%
\right)\eaa where
$\epsilon_1(k)=2(J_{1a}+J_{2a}-J_{1b}-J_{2b}+2J_3)+2J_{2b}\cos(k_x-k_y)$,
$\epsilon_2(k)=2J_{2a}\cos(k_x+k_y)+2J_3(\cos(2k_x)+\cos(2k_y))$ and
$f(k)=(e^{ik_x}+e^{ik_y})$. The explicit analytical expressions for the spin-wave dispersion
spectra are unreasonably  long and will not be given here. We
plot the 3-D spin-wave dispersion in Fig.[\ref{spinwave}].
Regardless of the values of parameters, a common feature of the
spin-wave dispersion is   an almost-dispersionless line along
$(k\pm\pi,k)$ where two branches become degenerate with an  energy
around
$4S\sqrt{(J_{1a}-J_{1b}+2J_{2a}-2J_{2b})(J_{1a}-J_{1b}-2J_{2b}+4J_3)}$.
By comparing the spin-wave dispersions for different $J_3$, we can
also determine the value of $J_3$. In Fig.[\ref{spinwave}], we
compare the dispersion along $(k,-k)$ for $J_3=0.1,0.4,0.8$
respectively. The major difference lies in the higher-branch,
which turns from a convex to a concave-shaped line.  This
feature can be tested in inelastic neutron scattering to determine
the value of $J_3$ explicitly.

 The spin moment correction in large $S$ limit is:
 \baa  \Delta S=\sum_{i=1,2}\int_{BZ}\langle
b_{i,k}^\dag b_{i,k}\rangle dk^2 \eaa
%\\
%\nonumber&=&\sum_{i=1,2}\int_{BZ}(|U_{i3}(k)|^2+|U_{i4}|^2)dk^2.\label{spinlength}\eaa
%$U$ is the Bogliubov transform matrix defined as\baa\left(%
%\begin{array}{c}
%  b_{1,k} \\
%  b_{2,k} \\
%  b^\dag_{1,-k} \\
%  b^\dag_{2,-k} \\
%\end{array}%
%\right)=U\left(%
%\begin{array}{c}
%  \alpha_{1,k} \\
%  \alpha_{2,k} \\
%  \alpha^\dag_{1,-k} \\
%  \alpha^\dag_{2,-k} \\
%\end{array}%
%\right)\eaa, where $\alpha$'s are the boson operator annihilating
%the eigenstates. Numerically but also exactly, we can calculate the correction in the AFM3 phase.
In the vicinity of AFM3-ICA transition line, $\Delta m$ is about
25\%. It is worth noting that $\Delta m$ does not diverge at the
phase boundary because the ICA (ICB) phases can be obtained from the
AFM3 state by a continuous rotation of the magnetization starting
from zero at the phase transition line, unlike in other phase
transitions where the magnetic order wavevector direction
changes\cite{Fang2008}.
 %For example, in the $J_1-J_2$ model at the critical
%point $J_1=2J_2$, the collinear magnetic order with $(\pi,0)$
%changes abruptly toanother  antiferromagnetic order with
%$(\pi,\pi)$,  the spinwave correction diverges\cite{Fang2008}.
%Another statement equivalent to the convergence of spinwave
%correction is that the density of states for spinwaves is finite
%everywhere in the BZ even at phase transitions to ICA or ICB. This
%can be easily inferred from the spinwave dispersion at the phase
%boundary. Taking the boundary between AFM3 and ICA, along (11)
%direction around (0,0) we have the quadratic dispersion\baa
%\omega(k,k)=\frac{4S(J_{2a}+2J_3)(J_{1a}+J_{2a}^t)k^2}{J_{1a}},\eaa
%but along $(1\bar{1})$ direction we have the linear dispersion as
%\baa\omega(k,-k)=4S\sqrt{(J_{1a}+J_{2a}^t)(J_{1a}-J_{1b}-4J_{2b}+8J_3)}|k|.\eaa

\begin{figure}[t]
\includegraphics[width=8cm, height=6cm]{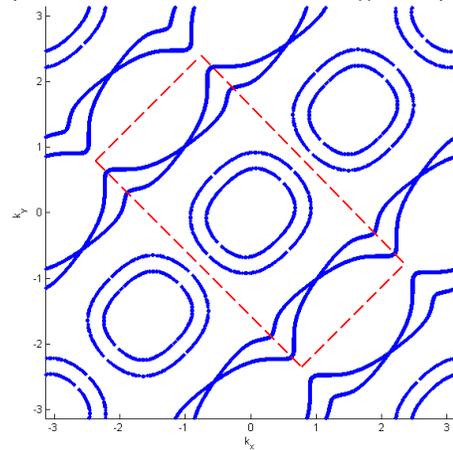}
\caption{The Fermi surface of the system in the magnetic ordered
AFM3 order taking $A_0=0.3t_1$. The dotted red lines crop out the
(folded) Brillion zone.
   \label{fermisurface}}
\end{figure}

We now discuss the influence of the observed magnetic order in the
AFM3 phase on the electronic properties of the material. DFT
calculations show that $FeSe$ and $FeTe$ have a very similar Fermi
surface structure to that in $Fe-As$ based materials. In the AFM3
state, the meanfield Hamiltonian can be written as $ H_{mf}= H_0+
H_M$, where $H_0=\sum_{ k\sigma}
\psi^+_{\sigma}(k)\epsilon(k)\psi_{\sigma}(k)$ is kinetic energy
that describes the band structure and $H_M$ is the mean-field energy
of the spin ordering,  \ba
H_M=\sum_{k,q}A(q)(\psi^+_{\uparrow}(k+q)\psi_{\uparrow}(k)-
\psi^+_{\downarrow}(k+q)\psi_{\downarrow}(k)) \ea with  $ A(q)=
A_0\sum_{j=0}^1 (1+i(-1)^j)\delta(q-q_j)$ where
$q_0=(\frac{\pi}{2},\frac{\pi}{2}),
q_1=(\frac{-\pi}{2},\frac{-\pi}{2})$ in the AFM3 state. The
resulting Fermi surface is given in Fig.[\ref{fermisurface}] and
remains gapless  even at a considerably large order parameter
$A_0=0.3t_1$. The reason is simple: the modulation vector is
$(\pi/2,\pi/2)$. This vector, and its multiples, can only couple two
electron or two hole pockets, but cannot couple electron and hole
pockets together. This is in sharp contrast to the 1111 system,
where the magnetic order induces a partial gap at both the electron
and hole Fermi pockets. This distinctive feature from 1111 or 122
systems may be detected using ARPES.

It is also important to discuss what happens upon replacing  $Te$
atoms by $Se$ atoms in the parent compound. Experimentally, it was
shown that superconductivity develops in $FeTe_{1-x}Se_{x}$ upon
increasing $x$. Based on our model, the AFM3 phase is generated from
strong coupling of lattice and magnetic degrees of freedom. The AFM3
phase must coexist with monoclinic lattice distortion which is not
favored in the pure $\alpha-FeSe$. Therefore, by increasing $Se$
concentration, the magnetic frustration increases. The frustration
can lead to a close competition between the AFM3 and AFM2 phases
which can result in a strong quantum critical region controlled by a
quantum critical point between the AFM3 and AFM2 phases or an
existence of a spin liquid state\cite{Ferrer1993}, which is an
interesting open question to study in future. This physics will be
critical to understanding unconventional transport properties in the
materials\cite{Wangnl} at high temperature or upon increasing the
$Se$ concentration.

In summary, we constructed a magnetic exchange model to explain the
rich magnetic order in $Fe_{1+y}Te_{1-x}Se_x$.  The model exhibits
both commensurate antiferromagnetic and incommensurate magnetic
order along the crystal a-axis and describes the transition from the
commensurate to the incommensurate spin order upon increasing the
concentration of excess $Fe$ atoms.  Our model can be explicitly
tested by experimentally measurable spin-wave features. Our model
also suggests an existence of a large critical region due to strong
spin frustration upon increasing $Se$ concentration.

 {\it Acknowledgments} JPH thanks S.
Kivelson, Pengcheng Dai, Igor Mazin, Tao Xiang, ZY Lu, Wei Bao,
Shiliang Li, H. Yao , W. Tsai, and DS Yao   for useful discussions.
BAB thanks P. W. Anderson and N. P. Ong for useful discussions. JPH
and CF were supported by the NSF under grant No. PHY-0603759.

\end{document}